\documentclass[a4paper,12pt]{jpconf}

\usepackage{graphicx}
\usepackage{bm}

\usepackage[square,comma,numbers,sort&compress]{natbib}

\usepackage{indentfirst}

\begin{document}

\title{Pseudorapidity dependence of multiplicity and transverse momentum fluctuations in pp collisions at SPS energies}

\author{Daria Prokhorova for the NA61/SHINE Collaboration}

\address{St. Petersburg State University, St. Petersburg, 199034, Russia}

\ead{daria.prokhorova@cern.ch}

\begin{abstract}

A search for the critical behavior of strongly interacting matter was performed at the NA61/SHINE experiment by studying event-by-event fluctuations of multiplicity and transverse momentum of charged hadrons produced in inelastic p+p collisions at 20, 31, 40, 80 and 158 GeV/c beam momentum. Results for the energy dependence of the scaled variance of the multiplicity distribution and for two families of strongly intensive measures of multiplicity and transverse momentum fluctuations $\Delta[P_{T},N]$ and $\Sigma[P_{T},N]$ are presented. These quantities were studied in different pseudorapidity intervals, which correspond to changing the baryon chemical potential and the temperature at the freeze-out stage. The strongly intensive measures $\Delta[N_{F},N_{B}]$ and $\Sigma[N_{F},N_{B}]$ were also used in the analysis of short- and long-range multiplicity correlations. Results on multiplicity and transverse momentum fluctuations significantly depend on the charges of the selected hadrons and the width and/or location of pseudorapidity intervals. The event generator EPOS does not describe the data for the $\Delta[P_{T},N]$ measure, but provides a fair description of $\Sigma[P_{T},N]$. The measure $\Sigma[N_{F},N_{B}]$ of forward-backward fluctuations is reproduced reasonably well by the EPOS model.

\end{abstract}

\section{Introduction}

Nowadays numerous experimental and theoretical investigations of high energy nucleus-nucleus collisions show that the quark-gluon plasma could exist in nature. Moreover, the results of CERN LHC \cite{LHSres1} and BNL RHIC \cite{RHICres} experiments and the observation of the transition between hadronic matter and quark-gluon plasma at CERN SPS energies \cite{SPSres11}, \cite{SPSres2}, \cite{SPSres3} revealed that the key question in nuclear and particle physics now is to determine the structure of the phase transition between the hadron gas and the deconfined matter. It is expected that in the phase diagram of strongly interacting matter the hadron gas and quark-gluon plasma regions are separated by a first order phase transition line at high baryo-chemical potentials and moderate temperatures. A crossover between both phases is assumed for high temperatures and low baryo-chemical potentials. The first order phase transition line then ends in a critical point. But the exact location of the critical end-point in the phase diagram is unknown. Moreover, some lattice QCD calculations suggest that there might be no critical point at all and only a crossover separates the two phases.

The strong interactions programme of NA61/SHINE, a fixed-target experiment at the CERN SPS \cite{layout}, includes the study of the properties of deconfinement and the search for the critical behaviour of strongly interacting matter. The main strategy of the NA61/SHINE collaboration in the search for the critical point is to perform a comprehensive two-dimensional scan of the phase diagram of strongly interacting matter by changing the collision energy and the system size \cite{SEARCH}. If the critical point exists it is expected that at some values of these parameters a region of increased fluctuations should be observed. At the top of this \textquotedblleft shark fin\textquotedblright\ hill the value of the critical fluctuations is expected to be a maximum. However, the critical signal could be shadowed since results on fluctuations are sensitive to conservation laws, resonance decays, volume fluctuations in the system of the colliding nuclei, quantum statistical effects and the limited acceptance of the experiment. Hence one should try to reduce the contribution from trivial fluctuations. This led to the idea to use intensive and strongly intensive quantities as probes of the critical behaviour \cite{ppSPS}, \cite{values}.

\section{Quantities of interest}

%\subsection{Intensive quantity}

In order to make proper comparison of the results from different colliding systems, one should choose so-called intensive variables which are independent of the system size. Since in the vicinity of a critical point central second moments of distributions of extensive event quantity $A$ are believed to diverge \cite{divergence1}, the scaled variance $\omega[A]$, an intensive quantity, was chosen for the analysis. The normalization results in $\omega[A]$ = 0 in the absence fluctuations of $A$ and  $\omega[A]$ = 1 in the case of a Poisson distribution of $A$ \cite{onnorm}. One should keep in mind that $\omega[A]$ is still sensitive to fluctuations of the volume.

%\subsection{Strongly intensive quantities}

Due to the imperfect centrality determination in A+A collisions, one should expect event-by-event volume fluctuations. Consequently, to eliminate the influence of usually poorly known distributions of the system volume, it was suggested to use strongly intensive quantities which are independent both of the volume and fluctuations of the volume within the statistical model of the ideal Boltzmann gas in the grand canonical ensemble formulation \cite{values}, \cite{onnorm}. Two families of strongly intensive variables $\Delta[A,B]$ and $\Sigma[A,B]$ were suggested which are functions of two extensive event quantities $A$ and $B$.

The normalization of these variables can be chosen such that~\cite{onnorm}:
\begin{itemize}
\item $\Delta[A,B]=\Sigma[A,B]=0$ in the absence of fluctuations of $A$ and $B$
\item $\Delta[A,B]=\Sigma[A,B]=1$ in the Independent Particle Model
\end{itemize}

\section{Analysis details}

The main goal of this work is to extend the investigation of the phase diagram by measuring the pseudorapidity dependence of fluctuations. Analysis of proton-proton collisions is the baseline for future investigations of nucleus-nucleus collisions.

This paper presents results referring to all charged hadrons with $p_{T}<1.5$ GeV/c produced in the analysis acceptance of the NA61/SHINE experiment \cite{accmap} in inelastic proton-proton collisions at beam momenta of 20, 31, 40, 80 and 158 GeV/c. Only data-based corrections for off-target interactions were performed. Simulation-based corrections using the event generator EPOS1.99 \cite{EPOS} and the NA61/SHINE detector simulation chain are in progress for other biases.

The analysis has two parts. In the one window analysis the intensive quantity $\omega[N]$ and the strongly intensive quantities $\Delta[P_{T},N]$ and $\Sigma[P_{T},N]$ are studied as functions of the width of the chosen pseudorapidity interval. This corresponds to changing the rapidity averaged baryo-chemical potential at the freeze-out stage \cite{eschpot}. The two windows analysis performs a search for short- and long-range correlations via study of the dependence of the strongly intensive quantity $\Sigma[N_{F},N_{B}]$ on the distance between two separated pseudorapidity windows in which the Forward and Backward multiplicities are evaluated.

\subsection{The one window analysis}

Here $A$ was taken as the event multiplicity of charged hadrons $N$ and $B$ as the scalar sum $P_{T}$ of their transverse momenta. With $\langle \cdots\rangle$ denoting the average value over all events and $\overline{\cdots}$ the inclusive average value (over all particles and all events), one can define the following intensive quantities \cite{ppSPS}:
\begin{equation}
\omega[N]=\frac{\langle N^2\rangle - \langle N\rangle^2}{\langle N\rangle}, \qquad \omega[P_{T}]=\frac{\langle P_{T}^2\rangle - \langle P_{T}\rangle^2}{\langle P_{T}\rangle}
\end{equation}
and strongly intensive quantities \cite{ppSPS}:
\begin{equation}
\Sigma[P_{T},N]=\frac{1}{C_\Delta}\left[\langle N\rangle\omega[P_{T}]+\langle P_{T}\rangle\omega[N]-2\cdot (\langle P_{T}\cdot N\rangle-\langle P_{T}\rangle\langle N\rangle)\right]
\end{equation}
\begin{equation}
\Delta[P_{T},N]=\frac{1}{C_\Delta}\left[\langle N\rangle\omega[P_{T}]-\langle P_{T}\rangle\omega[N]\right],
\end{equation}
with normalization \cite{onnorm}: 
\begin{equation}
C_\Delta=C_\Sigma=\langle N\rangle\omega(p_{T}),\qquad  \omega(p_{T})=\frac{\overline{p_{T}^2} - \overline{p_{T}}^2}{\overline{p_{T}}}    
\end{equation}

\subsection{The two windows analysis}

Taking the extensive event quantities $A$ as the multiplicity $N_{F}$ in the Forward pseudorapidity window and $B$ as the multiplicity $N_{B}$ in the Backward pseudorapidity window, the above formulae will transform \cite{Andronov} into :
\begin{equation}
\Sigma[N_{F},N_{B}]=\frac{1}{C_\Sigma}\left[\langle N_{B}\rangle\omega[N_{F}]+\langle N_{F}\rangle\omega[N_{B}]-2\cdot (\langle N_{F}\cdot N_{B}\rangle-\langle N_{F}\rangle\langle N_{B}\rangle)\right]
\end{equation}
\begin{equation}
\omega[N_{F}]=\frac{\langle N_{F}^2\rangle - \langle N_{F}\rangle^2}{\langle N_{F}\rangle}, \qquad \omega[N_{B}]=\frac{\langle N_{B}^2\rangle - \langle N_{B}\rangle^2}{\langle N_{B}\rangle}, \qquad C_\Sigma=\langle N_{B}\rangle+\langle N_{F}\rangle 
\label{omega}
\end{equation}

\subsection{Definitions of pseudorapidity intervals}

All results are presented as functions of $\Delta\eta/\Delta\eta_{max}$ in the lab system according to the Fig.1. In the following the lower edges of the pseudorapidity intervals are moving from $y_{beam}^{lab}/2$ to $y_{beam}^{lab}$ corresponding to the range from $y^{cms}=0$ to $y_{beam}^{cms}$. $\eta^{lab}$ is restricted to this range to exclude the influence of the bad acceptance coverage at small $\eta^{lab}$ and to reduce contributions from elastic processes at $\eta^{lab} > y_{beam}^{lab}$.
\begin{figure}[h!]
\begin{center}
\begin{minipage}[h]{0.35\linewidth}
\center{\includegraphics[width=1\linewidth]{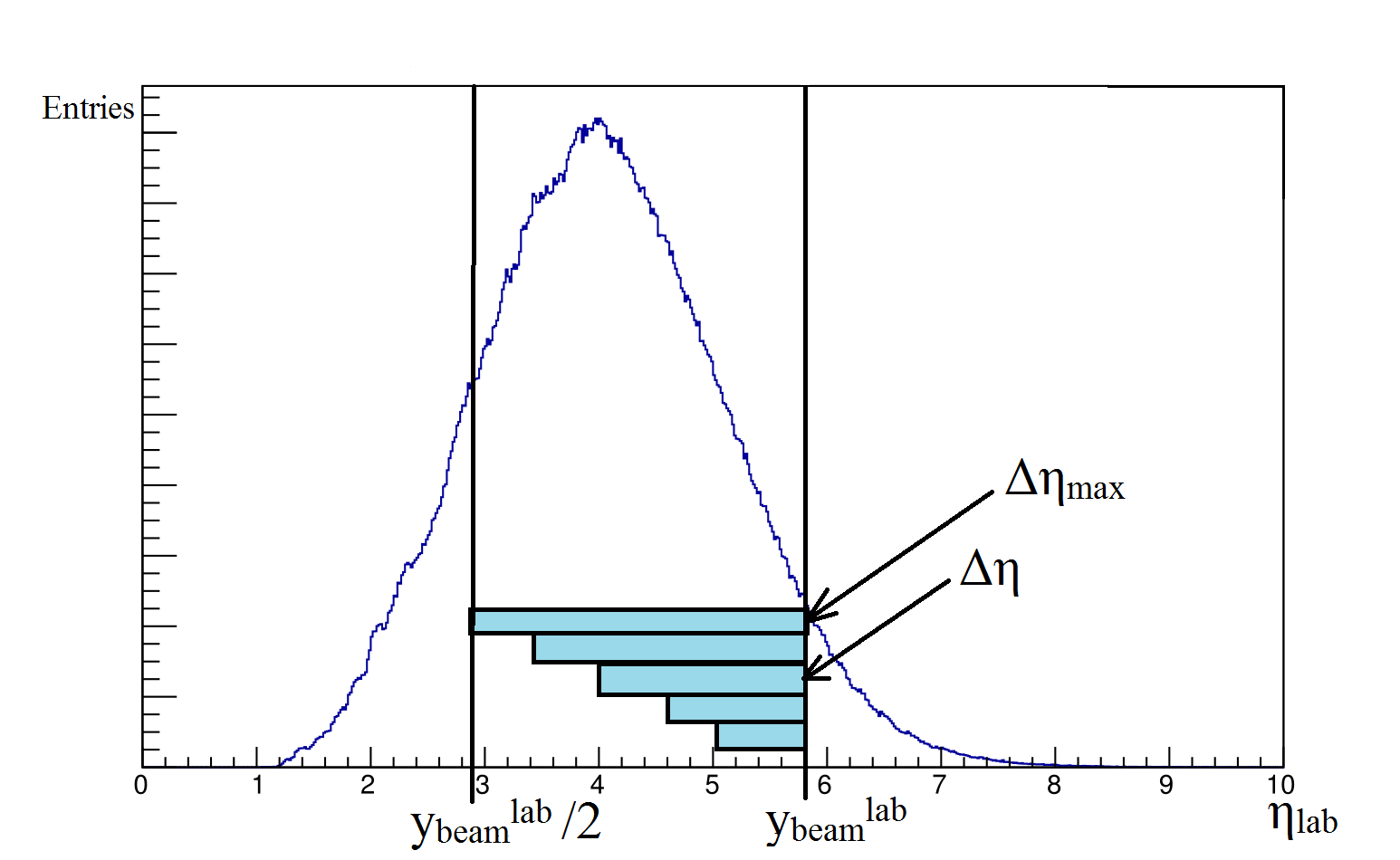}} \\a)\\
\label{etaa}
\end{minipage}
\qquad\quad
\begin{minipage}[h]{0.37\linewidth}
\center{\includegraphics[width=1\linewidth]{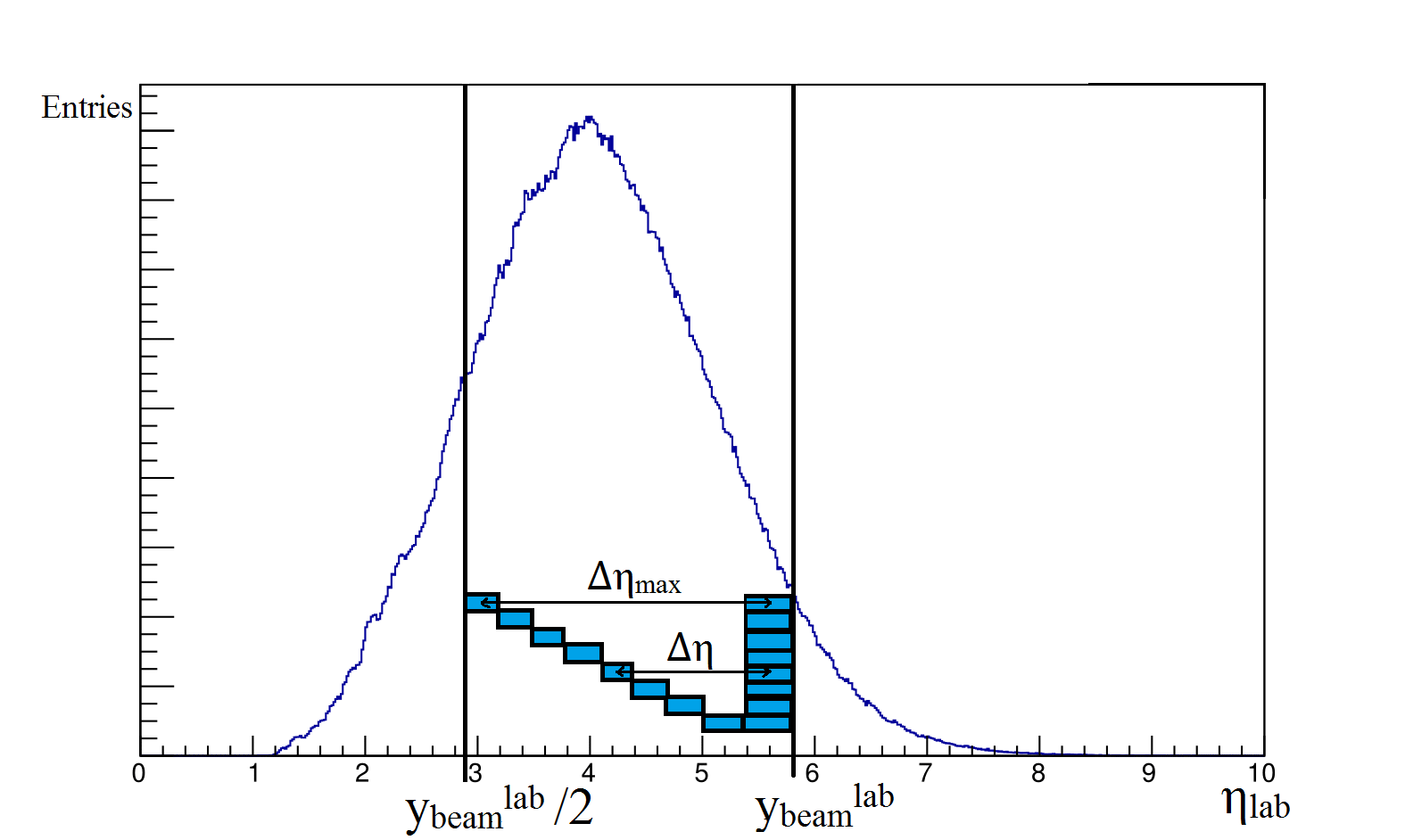}} \\b)\\ 
\label{etab}
\end{minipage}
\caption{Sketch of the uncorrected pseudorapidity distribution of charged hadrons with chosen windows (pseudorapidity intervals) a) for one window analysis, b) for two windows analysis}
\end{center}
\end{figure}
\section{Results}
The presented preliminary results refer to all charged hadrons with $p_{T}<1.5$ GeV/c produced in the acceptance of the NA61/SHINE experiment \cite{accmap} in inelastic p+p collisions.
\begin{figure}[h!]
\begin{center}
\begin{minipage}[h]{0.33\linewidth}
\center{\includegraphics[width=1\linewidth]{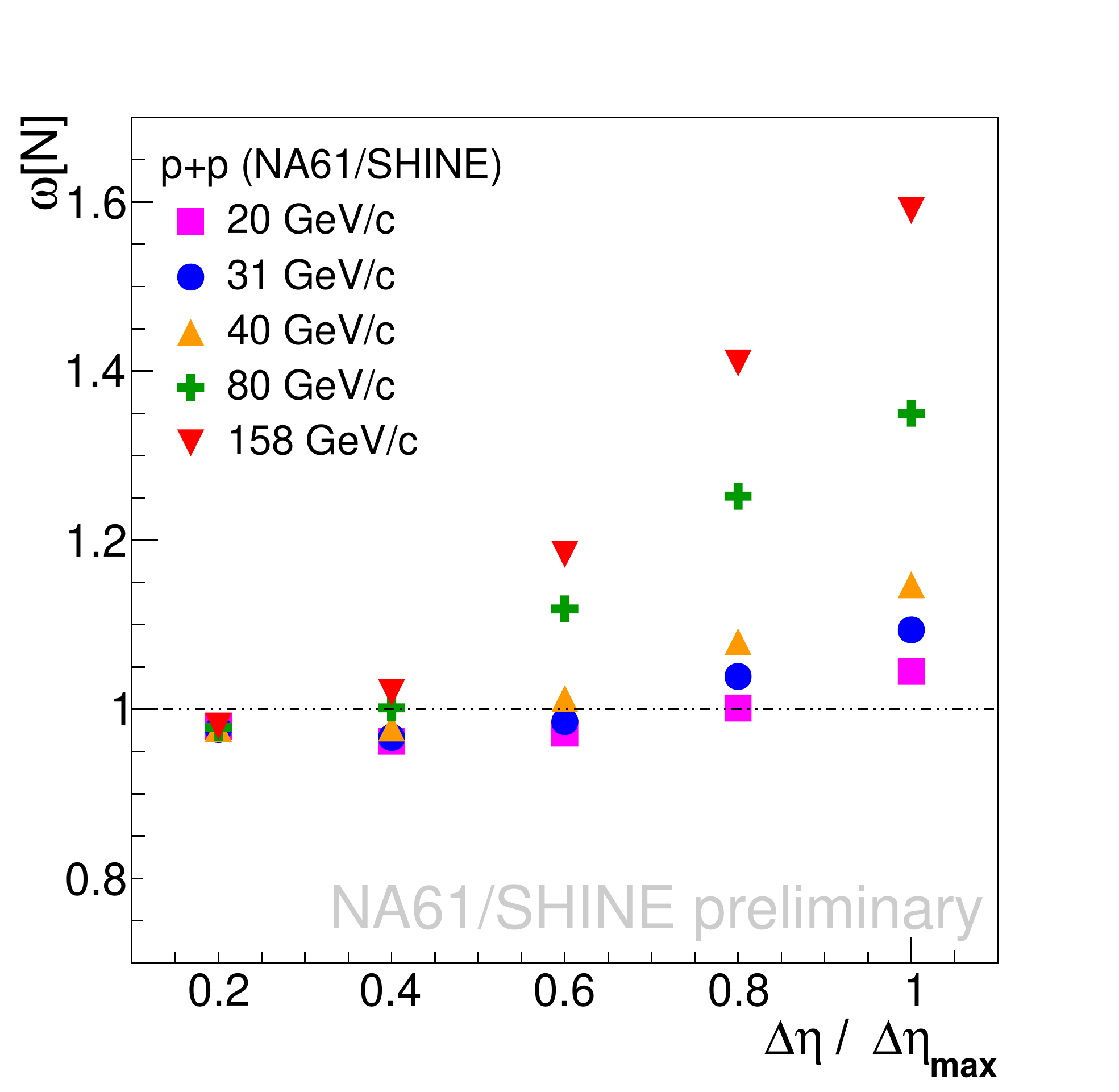}} \\a)\\
\end{minipage}
\qquad\quad
\begin{minipage}[h]{0.33\linewidth}
\center{\includegraphics[width=1\linewidth]{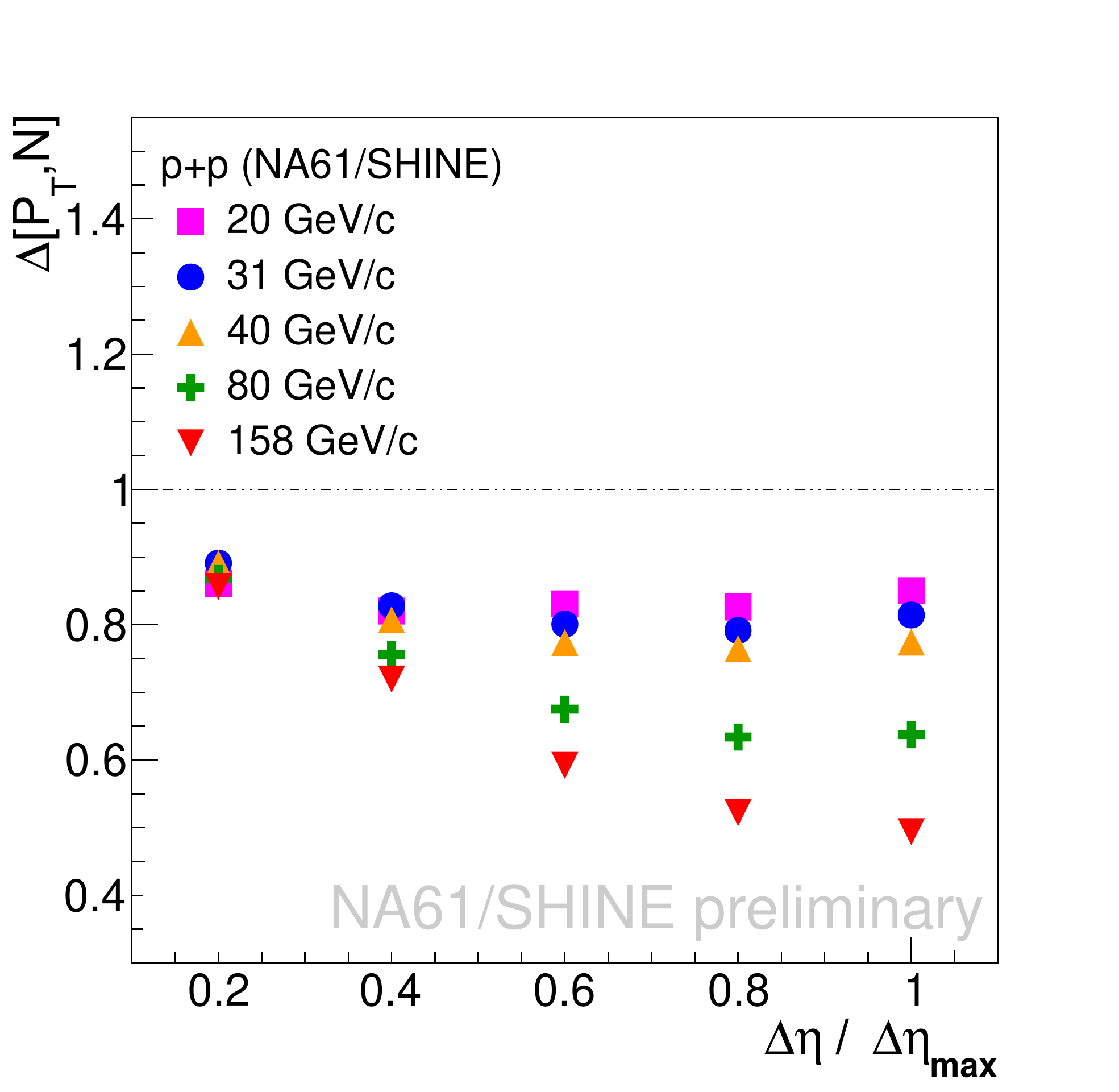}} \\b)\\ 
\end{minipage}
\caption{Dependence on the width of the rapidity window (see Fig.1a) of a) scaled variance $\omega[N]$, b) strongly intensive quantity $\Delta[P_{T},N]$ in the one window analysis at SPS energies for all charged hadrons produced in inelastic p+p collisions in the NA61/SHINE acceptance.}
\end{center}
\end{figure}
\begin{figure}[h!]
\begin{center}
\begin{minipage}[h]{0.33\linewidth}
\center{\includegraphics[width=1\linewidth]{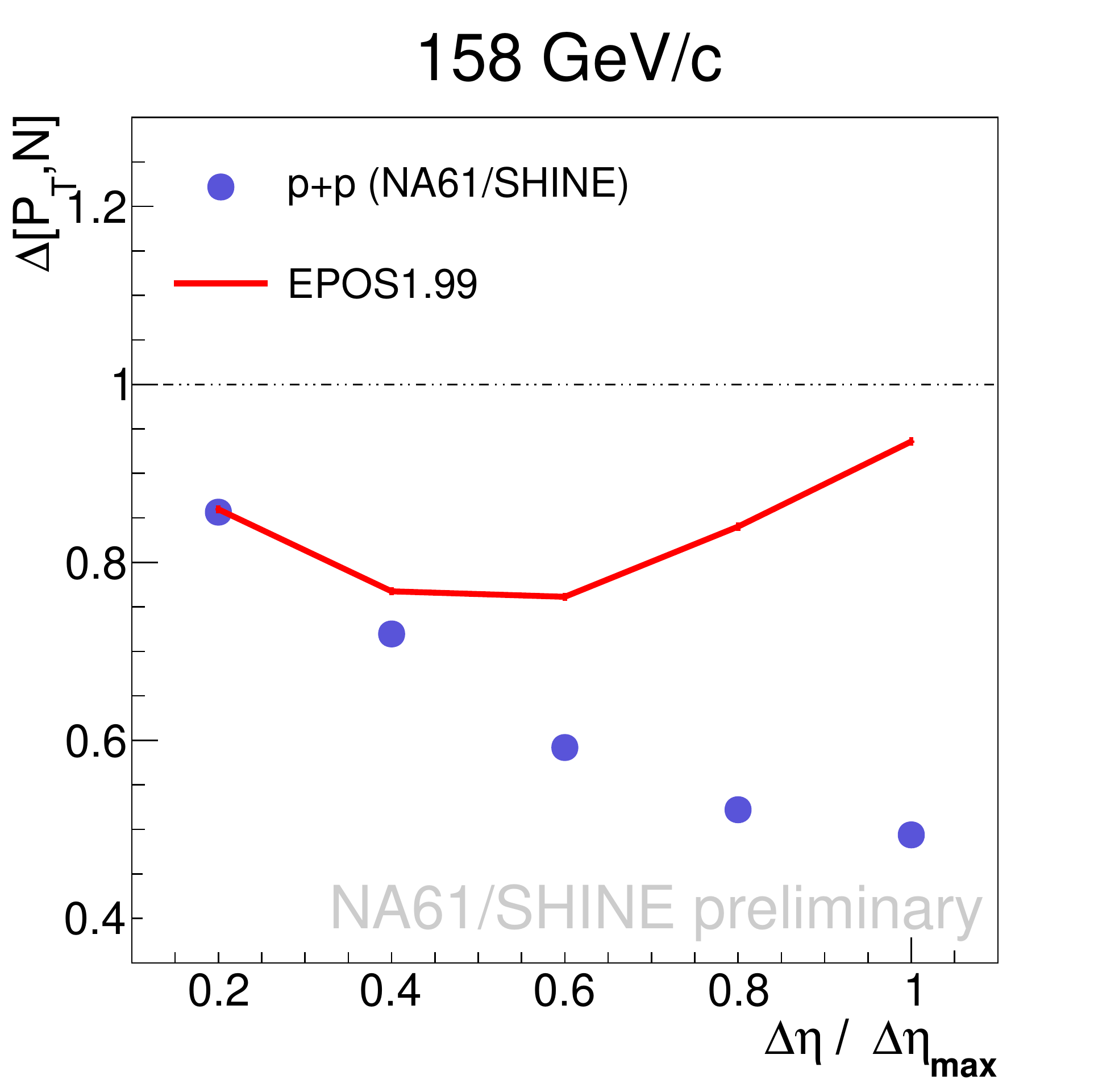}} \\a)\\
\end{minipage}
\qquad\quad
\begin{minipage}[h]{0.33\linewidth}
\center{\includegraphics[width=1\linewidth]{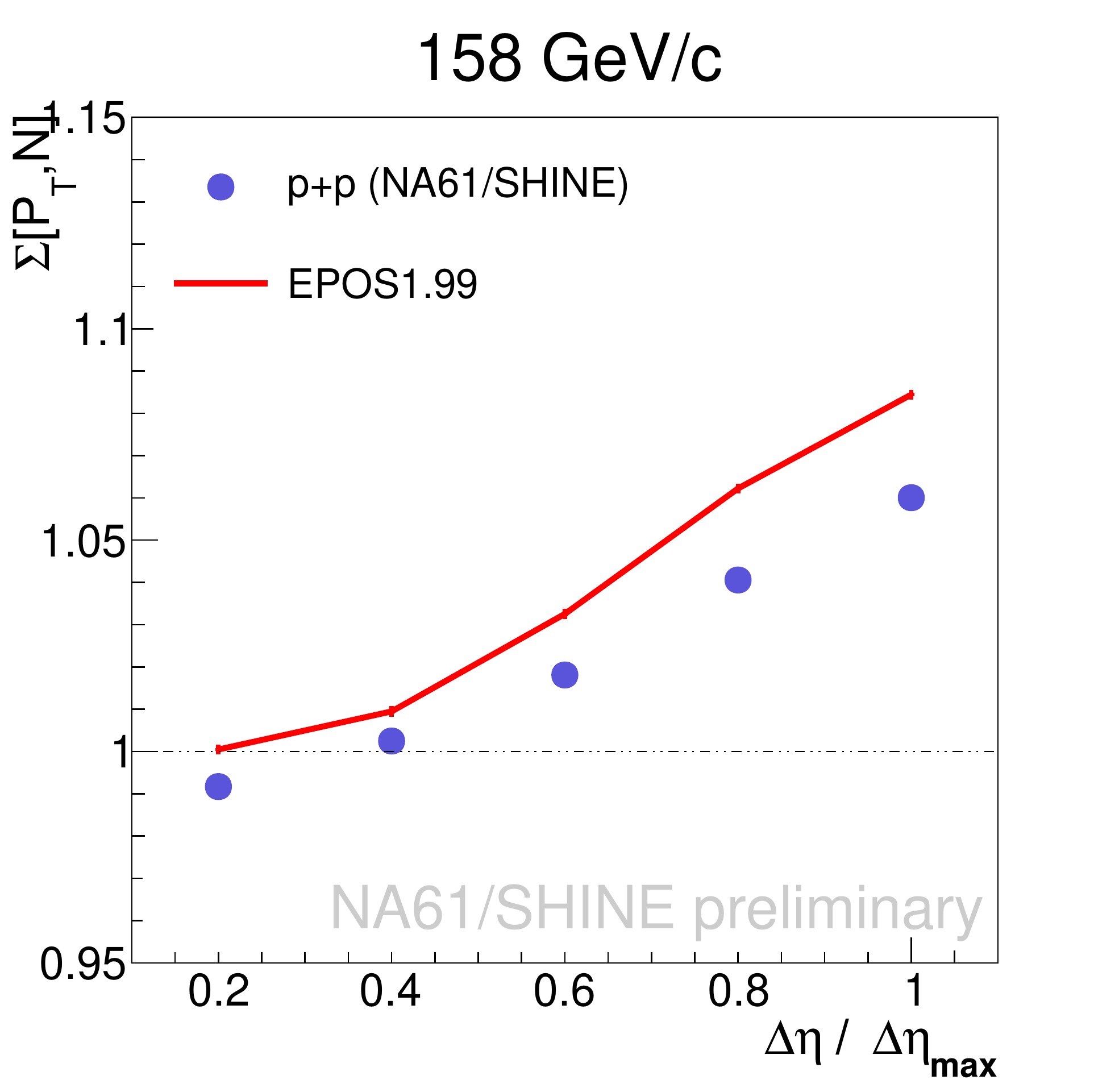}} \\b)\\ 
\end{minipage}
\caption{Dependence on the width of the pseudorapidity window (see Fig.1a) of strongly intensive quantities a) $\Delta[P_{T},N]$, b) $\Sigma[P_{T},N]$ in the one window analysis for data and EPOS1.99 at beam momentum of 158 GeV/c for all charged hadrons produced in inelastic p+p collisions in the NA61/SHINE acceptance.}
\end{center}
\end{figure}
\begin{figure}[h!]
\begin{center}
\begin{minipage}[h]{0.33\linewidth}
\center{\includegraphics[width=1\linewidth]{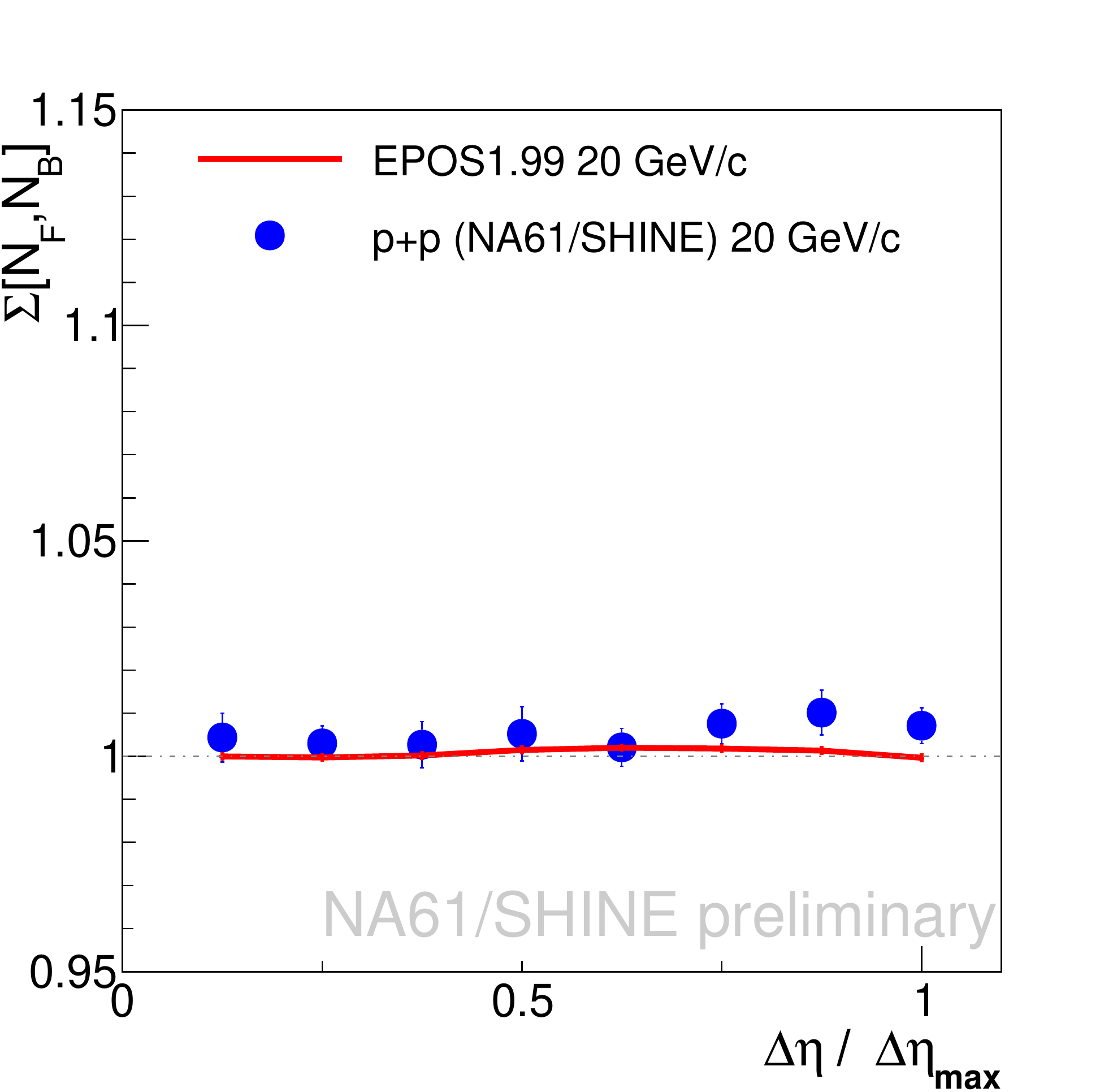}} \\a)\\
\end{minipage}
\qquad\quad
\begin{minipage}[h]{0.33\linewidth}
\center{\includegraphics[width=1\linewidth]{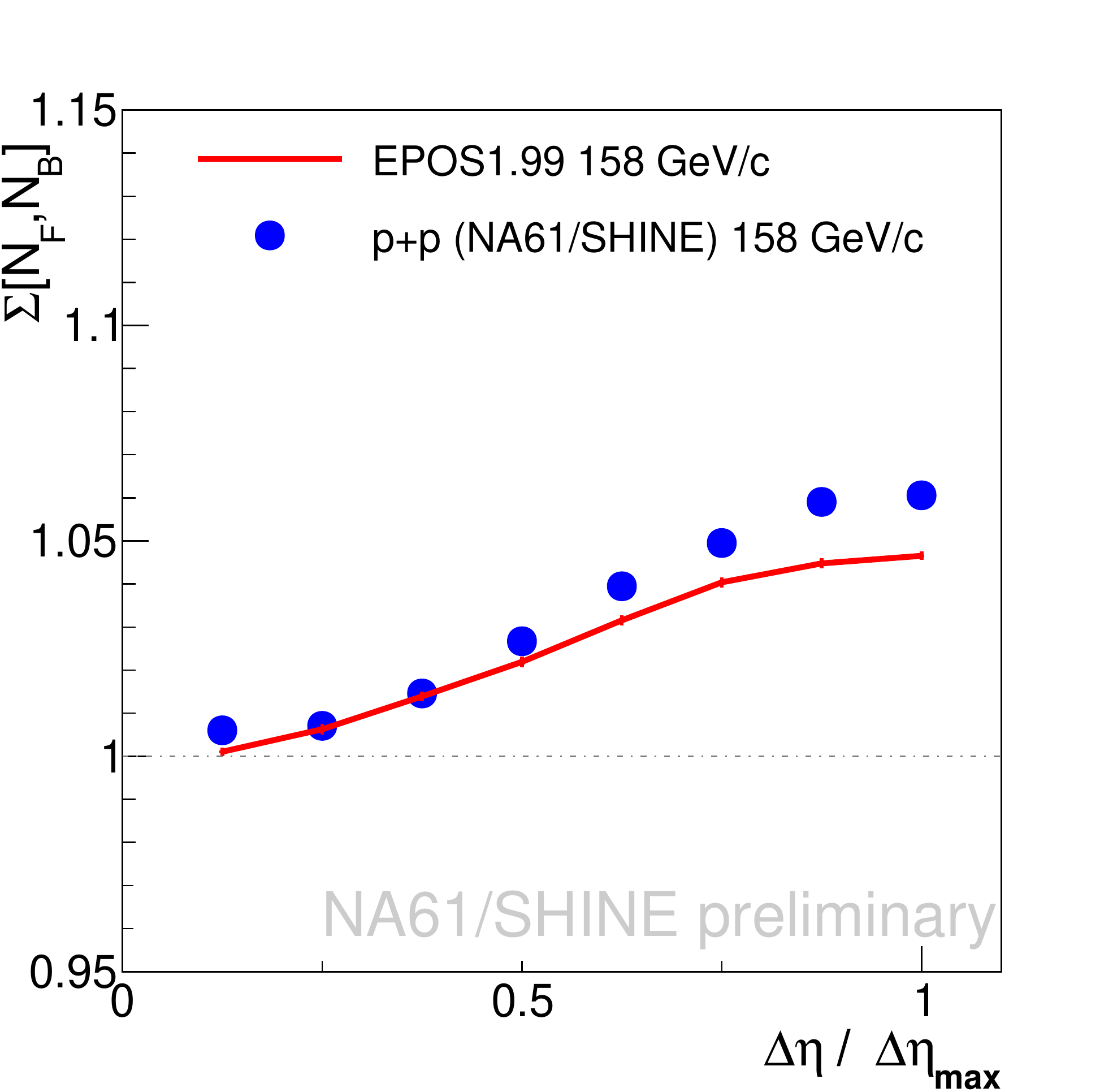}} \\b)\\ 
\end{minipage}
\caption{Dependence of the strongly intensive fluctuation measure $\Sigma[N_{F},N_{B}]$ for all charged hadrons produced in inelastic p+p collisions in the NA61/SHINE acceptance at beam momenta of a) 20 GeV/c, b) 158 GeV/c on the distance between the Forward and the Backward pseudorapidity window in the two windows analysis (see Fig.1b)}
\end{center}
\end{figure}

The studied fluctuation measures significantly depend on the width and the location of the pseudorapidity intervals. Results for $\omega[N]$ and $\Delta[P_{T},N]$ depend on the collision energy (see Fig.2). The values of $\omega[N]$ rise with increasing beam energy and width of the rapidity interval. The values of $\Delta[P_{T},N]$ are less than one with deviation increasing with beam energy and width of the rapidity interval. On the other hand the values of $\Sigma[P_{T},N]$ are similar at all beam momenta and show the same rising tendency with increasing width of the rapidity interval (see conference slides). The azimuthal acceptance changes only slightly for different beam momenta. A significant discrepancy between data and EPOS1.99 calculations is observed for $\Delta[P_{T},N]$ at all collision energies (see example for 158 GeV/c presented in Fig. 3, left). It is more pronounced for the larger width of the pseudorapidity interval. However, EPOS1.99 well describes $\omega[N]$ and $\Sigma[P_{T},N]$ (see Fig. 3, right and conference slides). In the two windows analysis there is an increase of the value of $\Sigma[N_{F},N_{B}]$ with the distance between forward and backward pseudorapidity intervals which is more pronounced for the higher energy (see Fig. 4). EPOS1.99 predictions are in a good agreement with the data.
%\newpage
\section *{Acknowledgments}
This work was supported by the Saint-Petersburg State University research grants 11.38.242.2015, 11.41.1385.2017.

\addcontentsline{toc}{section}{}

\bibliography{biblio}

\end{document}